# Title page

Names of the authors: F. Tárkányi[1], A. Hermanne[2], F. Ditrói[1], S. Takács[1]

Title: **Activation cross section data of proton induced nuclear reactions on lanthanum in the 34-65 MeV energy range and application for production of medical radionuclides**

Affiliation(s) and address(es) of the author(s):

[1] Institute for Nuclear Research, Hungarian Academy of Sciences (ATOMKI), Debrecen, Hungary

[2] Cyclotron Laboratory, Vrije Universiteit Brussel (VUB), Brussels, Belgium

E-mail address of the corresponding author: ditroi@atomki.hu

# Activation cross section data of proton induced nuclear reactions on lanthanum in the 34-65 MeV energy range and application for production of medical radionuclides

F. Tárkányi[1], A. Hermanne[2], F. Ditrói[1*], S. Takács[1]

[1] *Institute for Nuclear Research, Hungarian Academy of Sciences (ATOMKI), Debrecen, Hungary*

[2] *Cyclotron Laboratory, Vrije Universiteit Brussel (VUB), Brussels, Belgium*

## Abstract

Activation cross sections of the $^{nat}$La(p,x)$^{139,137m,137g,135,134,133m}$Ce, $^{nat}$La(p,x)$^{135,134,133}$La and $^{133m,133g,131}$Ba nuclear reactions have been measured experimentally, for the first time (except $^{139}$Ce). Cross-sections in the 34-64 MeV energy range were obtained through an activation method combining the stacked foil irradiation technique and gamma-ray spectrometry. The experimental cross sections were compared with the theoretical prediction in the TENDL-2014 and TENDL-2015, TALYS based libraries. Applications in the field of medical radionuclides production is discussed.

[*] Corresponding author: ditroi@atomki.hu

# Introduction

In the frame of our systematic study of light charged particle production routes of medical isotopes we have investigated the proton induced reactions on lanthanum targets. Among the large variety of possible reaction products the following radionuclides are presently used for applications, mainly in the medical research:

- The relatively long-lived $^{139}$Ce ($T_{1/2}$ = 137.6 d) [1] is useful as a standard for the calibration of γ-ray detectors and was proposed also for radiotherapy [2]
- $^{135}$La ($T_{1/2}$ = 19.5 h) can be used for Auger-based therapy [3, 4]
- Among a variety of cerium radioniclides, $^{137m}$Ce ($T_{½}$ = 34.4h, IT (99.22%), β+ (0.779%)) could be a novel candidate radionuclide in the field of diagnosis owing to its appropriate half-life, and its intense gamma-line at 254.29 keV [5].
- $^{134}$Ce/$^{134}$La ($T_{1/2}$ = 3.16 d / 6.45 min) an Auger electron and positron-emitting generator pair for radionuclide therapy [6]
- $^{131}$Cs ($T_{1/2}$ = 9.689 d) for brachytherapy treatment [7]

We earlier reported on production routes of these radioniclides through irradiation of Xe, Cs, Ba and Pr targets with protons, deuterons and alpha particles. [8, 9, 10, 11, 12, 13, 14, 15, 16]. Investigations of nuclear reactions on cerium for formation of these radionuclides is in progress.

# Experiment and data evaluation

Experimental determination of the cross-sections was done by an activation method combining the stacked foil irradiation technique and off-line gamma-ray spectrometry. The main experimental parameters and methods used in data evaluation for the present study are summarized in Table 1. The relevant decay data and the contributing reactions are presented in Table 2. The reproduction of the excitation function of the simultaneously irradiated $^{27}$Al(p,x)$^{24,22}$Na monitor reaction, showing good agreement with the recommended data, is shown in Fig. 1.

Table 1. Main experimental parameters and data evaluation

| Experimental parameters | | Data evaluation | |
|---|---|---|---|
| Incident particle | Proton | Gamma spectra evaluation | Genie 2000 [17] Forgamma[18] |
| Method | Stacked foil | Determination of beam intensity | Faraday cup (preliminary) Fitted monitor reaction (final)[19] |
| Target and thickness | La foil, 22.2μm | Decay data | NUDAT 2.6 [20] |
| Number of La target foils | 24 | Reaction Q-values | Q-value calculator [21] |
| Target composition and thickness (μm) | Tb(22.2), Al(377), La(25), Al(10), CeO(33.1-63.2, sedimented), Al (100) repeated 15 times | Determination of beam energy | Anderson (preliminary) [22] Fitted monitor reaction (final) [19] |
| Accelerator | Cyclone 90 cyclotron of the Université Catholique in Louvain la Neuve (LLN) | Uncertainty of energy | Cumulative effects of possible uncertainties |
| Primary energy (MeV) | 65 | Cross sections | Elemental cross section |
| Energy range (MeV) | 64.09 - 33.73 | Uncertainty of cross sections | Sum in quadrature of all individual linear contributions[23] |
| Irradiation time (min) | 59 | Yield | Physical yield [24, 25] |
| Beam current (nA) | 90 | | |
| Monitor reaction, [recommended values] | $^{27}$Al(p,x)$^{24}$Na reaction [26] | | |
| Monitor target and thickness (μm) | $^{nat}$Al, 377 and (100+10) | | |
| Detector | HPGe | | |
| γ-spectra measurements | 4 series | | |
| Cooling times (h) | 9.4-13.0;  30.1-49.8 240.1-342.4;  583.9-1539.5 | | |

Table 2 Decay and nuclear characteristic of the investigated reaction products [20], contributing reactions and their Q-values [21].

| Nuclide Spin Isomeric level | Half-life | Decay method (%) | $E_\gamma$ (keV) | $I_\gamma$ (%) | Contributing process | Q-value (keV) |
|---|---|---|---|---|---|---|
| $^{139}$Ce 3/2+ | 137.64 d | ε: 100 | 165.8575 | 80 | $^{139}$La(p,n) | -1060.73 |
| $^{137m}$Ce 11/2- 254.29 keV | 34.4 h | IT: 99.21 ε: 0.79 | 254.29 | 11.1 | $^{138}$La(p,2n) $^{139}$La(p,3n) | -9456.76 -18234.77 |
| $^{137g}$Ce 3/2+ | 9.0 h | ε: 100 | 436.59 447.15 | 0.250 1.68 | $^{138}$La(p,2n) $^{139}$La(p,3n) $^{137m}$Ce decay | -9456.76 -18234.77 |
| $^{135}$Ce 1/2(+) | 17.7 h | ε:100 | 265.56 300.07 518.05 572.26 606.76 783.59 | 41.8 23.5 13.6 10.4 18.8 10.6 | $^{138}$La(p,4n) $^{139}$La(p,5n) | -26901.9 -35679.9 |
| $^{134}$Ce 0+ | 3.16 d | ε:100 | 130.414 162.306 | 0.209 0.230 | $^{138}$La(p,5n) $^{139}$La(p,6n) | -34756.6 -43534.6 |
| $^{133m}$Ce 9/2- 37.27 keV | 5.1 h | ε: 100 | 130.803 477.22 | 18.0 39.3 | $^{138}$La(p,6n) $^{139}$La(p,7n) | -45242.6 -54020.6 |
| $^{135}$La 5/2+ | 19.5 h | ε: 100 | 480.51 874.51 | 1.52 0.16 | $^{138}$La(p,p3n) $^{139}$La(p,p4n) | -24092.4 -32870.36 |
| $^{134}$La 1+ | 6.45 min | ε: 100 β$^+$: 63.6 | 563.246 604.721 | 0.362 5.04 | $^{138}$La(p,p4n) $^{139}$La(p,p5n) | -33588.5 -42366.5 |
| $^{133}$La | | | | | $^{138}$La(p,p5n) $^{139}$La(p,p6n) | -41384.1 -50162.1 |
| $^{133m}$Ba 11/2- 288.25 keV | 38.93 h | IT: 99.9896 | 275.925 | 17.69 | $^{138}$La(p,2p4n) $^{139}$La(p,2p5n) $^{133}$La decay | -38542.56 -47320.57 |
| $^{133g}$Ba 1/2+ | 10.551 y | ε: 100 | 80.9979 276.3989 302.8508 356.0129 383.8485 | 32.9 7.16 18.34 62.05 8.94 | $^{138}$La(p,2p4n) $^{139}$La(p,2p5n) $^{133m}$Ba decay $^{133}$La decay | -38542.56 -47320.57 |
| $^{131}$Ba 1/2+ | 11.50 d | ε: 100 | 123.804 216.088 373.256 496.321 | 29.8 20.4 14.40 48.0 | $^{138}$La(p,2p6n) $^{139}$La(p,2p7n) $^{133}$La decay | -55554.86 -64332.86 |

The Q-values refer to formation of the ground state. In case of formation of a higher laying isomeric state it should be corrected with the energy of level energy of the isomeric state shown in Table 2.

When complex particles are emitted instead of individual protons and neutrons the Q-values have to be decreased by the respective binding energies (pn→d +2.2 MeV, p2n→t +8.5 MeV, 2pn→$^3$He +7.7 MeV, 2p2n→α +28.3 MeV).

Isotopic abundances: $^{138}$La ( 0.0902 % ), $^{139}$La ( 99.9098 %)

**Nuclear reaction model calculation**

As lanthanum is practically monoisotopic (see Table 1) the measured experimental data actually represent isotopic cross-sections, which can be directly compared to the model calculations. We use the theoretical data presented in the TENDL-2014 and TENDL-2015 [27] libraries, based on both default and adjusted TALYS (1.6) calculations [28], for the comparison in this study.

**Results**

*Cross sections*

Data are displayed in Figs. 2–12 together with the only earlier published result and with the TALYS predictions. The numerical values with uncertainties are given in Table 3 and Table 4. Although the cross-sections are in principle elemental, La can be practically considered as monoisotopic ($^{139}$La abundance is 99.9098 %). Activities of different radioproducts were assessed from spectra measured at increasing cooling time after EOB, in order to comply with their half-lives and the decay properties.

Table 3. Activation cross sections for the $^{nat}$La(p, x)$^{139,137m,137g,135,134,133m}$Ce reactions

| E (MEV) | ΔE (MEV) | $^{139}$Ce σ (mb) | Δσ (mb) | $^{137m}$Ce σ (mb) | Δσ (mb) | $^{137g}$Ce σ (mb) | Δσ (mb) | $^{135}$Ce σ (mb) | Δσ (mb) | $^{134}$Ce σ (mb) | Δσ (mb) | $^{133m}$Ce σ (mb) | Δσ (mb) |
|---|---|---|---|---|---|---|---|---|---|---|---|---|---|
| 64.09 | 0.20 | 2.39 | 1.40 | 51.70 | 5.91 | 14.46 | 3.04 | 211.40 | 23.74 | 271.51 | 36.52 | 2.00 | 0.29 |
| 62.02 | 0.23 | 7.27 | 1.32 | 57.59 | 6.54 | 20.14 | 4.30 | 250.27 | 28.10 | 256.28 | 34.21 | 0.30 | 0.12 |
| 59.80 | 0.26 | 3.13 | 2.16 | 63.11 | 7.19 | 13.50 | 2.74 | 324.76 | 36.46 | 221.18 | 33.55 | | |
| 57.50 | 0.30 | 8.02 | 1.51 | 64.13 | 7.22 | 21.83 | 4.57 | 359.06 | 40.31 | 144.81 | 27.01 | | |
| 55.83 | 0.32 | 11.23 | 2.04 | 69.68 | 7.91 | 12.64 | 2.52 | 394.55 | 44.30 | 117.40 | 22.54 | | |
| 54.76 | 0.34 | 10.16 | 1.92 | 72.47 | 8.23 | 30.49 | 6.20 | 403.24 | 45.27 | 71.09 | 13.54 | | |
| 53.67 | 0.36 | 10.94 | 1.86 | 70.03 | 7.89 | 16.47 | 2.37 | 384.16 | 43.13 | 55.20 | 10.85 | | |
| 52.15 | 0.38 | 8.38 | 4.49 | 77.44 | 8.81 | 31.13 | 5.43 | 398.37 | 44.72 | | | | |
| 50.81 | 0.40 | 14.52 | 2.22 | 78.62 | 8.91 | 17.08 | 3.36 | 367.24 | 41.23 | | | | |
| 49.57 | 0.42 | | | 80.68 | 9.15 | 32.47 | 5.64 | 327.23 | 36.74 | | | | |
| 48.39 | 0.44 | 14.30 | 2.08 | 89.21 | 10.08 | 23.33 | 4.08 | 302.69 | 33.99 | | | | |
| 47.05 | 0.46 | | | 90.24 | 10.20 | 16.39 | 2.99 | 240.93 | 27.05 | | | | |
| 45.74 | 0.48 | 16.69 | 2.57 | 104.10 | 11.74 | 29.07 | 4.29 | 204.70 | 22.98 | | | | |
| 44.94 | 0.49 | 12.57 | 2.17 | 109.23 | 12.31 | 22.70 | 3.27 | 168.96 | 18.97 | | | | |
| 43.73 | 0.51 | 17.02 | 2.17 | 124.71 | 14.03 | 31.38 | 4.09 | 121.68 | 13.67 | | | | |
| 42.90 | 0.52 | 15.11 | 2.31 | 125.41 | 14.11 | 30.24 | 3.85 | 83.67 | 9.40 | | | | |
| 41.66 | 0.54 | 13.35 | 1.86 | 102.59 | 11.54 | 47.71 | 5.73 | 32.53 | 3.66 | | | | |
| 40.80 | 0.55 | 15.72 | 2.07 | 119.14 | 13.39 | 42.97 | 5.08 | 16.17 | 1.82 | | | | |
| 39.45 | 0.57 | 15.43 | 1.85 | 153.69 | 17.27 | 31.74 | 3.71 | 7.52 | 0.85 | | | | |
| 38.55 | 0.58 | 17.09 | 2.09 | 174.93 | 19.65 | 24.43 | 2.85 | 3.28 | 0.37 | | | | |
| 37.19 | 0.60 | 23.38 | 2.82 | 246.26 | 27.66 | 39.92 | 4.59 | 0.94 | 0.11 | | | | |
| 36.24 | 0.62 | 19.24 | 2.16 | 265.16 | 29.76 | 51.03 | 5.86 | 0.50 | 0.06 | | | | |
| 34.74 | 0.64 | 21.50 | 2.70 | 316.70 | 35.57 | 61.80 | 7.07 | 0.45 | 0.07 | | | | |
| 33.73 | 0.66 | 17.68 | 2.18 | 350.30 | 39.34 | 57.55 | 6.49 | 0.28 | 0.04 | | | | |

Table 4 . Activation cross sections for the $^{nat}$La (p, x)$^{135,134,133}$La, $^{133m,133g131}$Ba reactions

| E (MEV) | ΔE (MEV) | $^{135}$La σ (mb) | Δσ (mb) | $^{134}$La σ (mb) | Δσ (mb) | $^{133}$La σ (mb) | Δσ (mb) | $^{133m}$Ba σ (mb) | Δσ (mb) | $^{133g}$Ba σ (mb) | Δσ (mb) | $^{131}$Ba σ (mb) | Δσ (mb) |
|---|---|---|---|---|---|---|---|---|---|---|---|---|---|
| 64.09 | 0.20 | 78.30 | 9.17 | 5.73 | 0.90 | 20.75 | 3.66 | 5.73 | 0.90 | 14.26 | 1.74 | 7.83 | 0.89 |
| 62.02 | 0.23 | 80.26 | 9.42 | 7.58 | 1.01 | 7.52 | 2.06 | 7.58 | 1.01 | 11.55 | 1.36 | 4.43 | 0.52 |
| 59.80 | 0.26 | 89.15 | 10.37 | 7.32 | 0.98 | | | 7.32 | 0.98 | 12.09 | 1.46 | 2.01 | 0.25 |
| 57.50 | 0.30 | 65.85 | 7.64 | 7.42 | 0.99 | | | 7.42 | 0.99 | 10.77 | 1.31 | 0.62 | 0.09 |
| 55.83 | 0.32 | 66.15 | 7.65 | 8.71 | 1.17 | | | 8.71 | 1.17 | 12.56 | 1.54 | 0.26 | 0.07 |
| 54.76 | 0.34 | 65.98 | 7.68 | 10.15 | 1.31 | | | 10.15 | 1.31 | 13.28 | 1.58 | | |
| 53.67 | 0.36 | 46.97 | 5.47 | 8.95 | 1.16 | | | 8.95 | 1.16 | 9.65 | 1.36 | | |
| 52.15 | 0.38 | 31.59 | 3.70 | 10.23 | 1.44 | | | 10.23 | 1.44 | 14.60 | 1.72 | | |
| 50.81 | 0.40 | 28.17 | 3.31 | 10.03 | 1.63 | | | 10.03 | 1.63 | 12.48 | 1.65 | | |
| 49.57 | 0.42 | 27.96 | 3.29 | 9.80 | 1.34 | | | 9.80 | 1.34 | 9.28 | 2.14 | | |
| 48.39 | 0.44 | 21.16 | 2.51 | 9.56 | 1.18 | | | 9.56 | 1.18 | 14.82 | 1.87 | | |
| 47.05 | 0.46 | 24.63 | 2.95 | 8.78 | 1.17 | | | 8.78 | 1.17 | 16.54 | 2.35 | | |
| 45.74 | 0.48 | 16.33 | 1.95 | 8.52 | 1.07 | | | 8.52 | 1.07 | 14.06 | 1.85 | | |
| 44.94 | 0.49 | | | 7.04 | 0.90 | | | 7.04 | 0.90 | 10.39 | 1.27 | | |
| 43.73 | 0.51 | | | 6.91 | 0.86 | | | 6.91 | 0.86 | 9.57 | 1.21 | | |
| 42.90 | 0.52 | | | 5.36 | 0.72 | | | 5.36 | 0.72 | 8.16 | 1.10 | | |
| 41.66 | 0.54 | | | 4.15 | 0.56 | | | 4.15 | 0.56 | | | | |
| 40.80 | 0.55 | 1.74 | 0.38 | 3.02 | 0.40 | | | 3.02 | 0.40 | 4.16 | 0.67 | | |
| 39.45 | 0.57 | | | 2.55 | 0.32 | | | 2.55 | 0.32 | | | | |
| 38.55 | 0.58 | | | 1.83 | 0.25 | | | 1.83 | 0.25 | | | | |
| 37.19 | 0.60 | | | 1.19 | 0.20 | | | 1.19 | 0.20 | | | | |
| 36.24 | 0.62 | | | 0.61 | 0.07 | | | 0.61 | 0.07 | | | | |
| 34.74 | 0.64 | | | 0.11 | 0.10 | | | 0.11 | 0.10 | | | | |
| 33.73 | 0.66 | | | | | | | | | | | | |

*Cerium radioniclides*

The Ce-radioniclides are formed by direct (p,xn) reactions. The production cross sections of long-lived ground states include the contribution from short-lived metastable state.

*Production of $^{139}$Ce*

The $^{139}$Ce ($T_{1/2}$ = 137.641 d) is produced via $^{139}$Ce(p,n) reaction. One earlier experimental data set is available in the literature by Vermeulen et al. [1] up to 20 MeV. We measured only the high energy tail of the excitation function (Fig. 2). It includes the complete decay of the short lived isomeric state ($T_{1/2}$ = 54.8 s, IT: 100 %) that corresponds rather well with the TALYS predictions.

*Production of $^{137m}$Ce*

The excitation function of the $^{137m}$Ce ($T_{1/2}$ = 34.4 h) isomeric state shown in Fig. 3 agrees well with the two, nearly identical, TENDL libraries.

*Production of $^{137g}$Ce*

The independent production cross-sections of $^{137g}$Ce ($T_{1/2}$ = 9.0 h) were obtained by subtracting the contribution of the decay of $^{137m}$Ce ( $T_{1/2}$ = 34.4 h, IT: 99.21 %) at the moment of measurement (Fig. 4). A slight upward energy shift is noted with respect to the TALYS predictions that are also shifted between themselves.

*Production of $^{135}$Ce*

The radionuclide $^{135}$Ce has two longer lived metastable states, the higher laying shorter-lived isomeric state ($T_{1/2}$ = 20 s, IT: 100 %) and the long-lived ground state ($T_{1/2}$ = 17.7 h, ε: 100 %). The measured cross section (Fig. 5) contains the production through the isomeric state decay (m+) and the direct production. The same energy shift between experimental data and the TALYS predictions as above is noted.

*Production of $^{134}$Ce*

The excitation function of the $^{134}$Ce (T$_{1/2}$ = 3.16 d, ε: 100 %) is shown in Fig. 6, showing good agreement with TENDL-2014 results. TENDL-2015 is higher and shifted by 5 MeV. The production was accessed through the weak (130.414 keV, I$_\gamma$= 0.209 %) gamma-line and through the gamma-lines of the short half-life daughter $^{134}$La (T$_{1/2}$ = 6.45 min, see Table 2).

*Production of $^{133m}$Ce*

Due to the long cooling time before first measurement we could obtain production cross sections (2 data points) only for the longer-lived isomeric state ($^{133m}$Ce T$_{1/2}$ = 5.1 h, ε: 100 %, $^{133g}$Ce T$_{1/2}$ = 97 min, ε: 100 %) (Fig. 7). Large difference between the two TENDL libraries is noted.

*Lanthanum radioniclides*

*Production of $^{135}$La*

The radionuclide $^{135}$La (T$_{1/2}$ = 19.5 h) is produced directly and through the decay of parent $^{135}$Ce (T$_{1/2}$ = 17.7 h, ε: 100 %). The independent production cross sections were obtained after subtracting the $^{135}$Ce decay contribution. The excitation function is shown in Fig. 8 and is in reasonable agreement with the two similar TENDL predictions.

*Production of $^{133}$La*

We obtained two cross-section points for production of $^{133}$La (T$_{1/2}$ = 3.912 h) near the effective threshold of the excitation function (Fig. 9). The $^{133}$La is produced directly and through the decay of the isomeric states of $^{133}$Ce ($^{133g}$Ce, T$_{1/2}$ = 97 min, ε: 100 % and $^{133m}$Ce, T$_{1/2}$ = 5.1 h, ε: 100 %). The measured cross-section contain the contribution of the shorter-lived $^{133}$Ce parent ground state. The contribution from the decay of the longer-lived isomeric state of the parent $^{133}$Ce was subtracted based on the results of theory but the correction at time of measurement was negligible taking into account the small cross section for production of $^{133m}$Ce (2 mb at 65 MeV, see Fig. 7).

*Barium radioniclides*

*Production of $^{133m}$Ba*

The radionuclide $^{133}$Ba has two longer-lived isomeric states. Out of them the $^{133m}$Ba isomeric state ($T_{1/2}$ = 38.93 h) is produced directly and through the decay of the $^{133}$La parent ($T_{1/2}$ = 3.912 h, $\varepsilon$ : 0.03 %). The cumulative cross section of $^{133m}$Ba is presented in the Fig. 10. The cross sections are overestimated in the TENDL-2014 library but are near the experimental values in TENDL-2015.

*Production of $^{133g}$Ba*

The long-lived ground state of $^{133}$Ba ($T_{1/2}$ = 10.551 y) is produced directly and through the decay of the $^{133m}$Ba isomeric state (38.93 h, IT: 99.9896) and of $^{133}$La ($T_{1/2}$ = 3.912 h, $\varepsilon$: 99.97.03 %) . The excitation function, showing some scatter due to the low count rates, is seen in Fig. 11. Large difference is shown in behavior at higher energy between the two TENDL predictions.

*Production of $^{131}$Ba*

The measured cross-sections of the $^{131}$Ba ground state ($T_{1/2}$ = 11.50 d) contain the contribution from the shorter-lived isomeric state (( $T_{1/2}$ = 14.6 min, IT 100 %) and of the $^{131}$La ($T_{1/2}$ = 59 min, $\varepsilon$: 100 %) decay. We could get data only around the effective threshold (Fig. 12) that are in good agreement with the TALYS predictions.

## Integral yields

The so called physical integral yields were calculated from a SPLINE fit of our experimental excitation functions and using fitted TALYS data in the missing energy range. The results are presented in Fig. 13 (except four activation products where we had only a few data points). No experimental thick target yields were published for the presently investigated energy range.

## Discussion of the production routes for radioniclides relevant for applications.

In the strict sense, production routes for a particular radionuclide can be only compared, when experimental data for all involved routes are available. It was, however, shown that reliable comparison is possible using the results of the TENDL library. The main advantage is that results are available for reactions induced by different charged particles on all target isotopes (with possibility to combine the results to elemental targets) and the agreement with the experimental data is acceptable for discussion and drawing conclusions. Many contributing factors should be taken into account in the comparison: production yield, radionuclidic purity, specific activity, required particle, required energy range, availability and price of the target material, complexity of the chemical separation, target recovery, target preparation, target heat conductivity and resistivity, target chemical form, etc. We will only discuss a few major factors here, mostly related to the target, irradiation parameters and radionuclidic purity.

*Production of $^{139}$Ce (T$_{1/2}$ = 137.6 d)*

The main light charged particle reaction routes for production of $^{139}$Ce include the following reactions: $^{139}$La(p,n), $^{138}$La(d,n), $^{139}$La(d,2n), $^{136}$Ba($\alpha$,n), $^{137}$Ba($\alpha$,2n), $^{138}$Ba ($\alpha$,3n), $^{141}$Pr(p,x) and $^{141}$Pr(d,x). The theoretical excitation functions are presented in Fig. 14. At low energies $^{139}$La(p,n) is the favorite reaction while at higher energies $^{139}$La(d,2n) and $^{141}$Pr(p,x) are preferred. A $^{nat}$La target can be used (ca 100 % $^{139}$La). The (d,2n) reaction has a higher yield than (p,n). The $^{141}$Pr(p,x) reaction has high yield and advantage of a naturally monoisotopic target. The alpha induced reactions require enriched targets and the yield is lower due to higher stopping.

*Production of $^{137m}$Ce (T$_{½}$ = 34.4 h)*

The main light charged particle reaction routes for production of $^{137m}$Ce include: $^{138}$La(p,2n), $^{139}$La(p,3n), $^{139}$La(d,4n), $^{138}$La(d,3n), $^{134}$Ba($\alpha$,n) and $^{135}$Ba($\alpha$,2n) (Fig. 15).
The $^{138}$La(p,2n) and the $^{138}$La(d,3n) are the most productive, but for the deuteron route long-lived $^{139}$Ce is co-produced, with an activity that is about 5000 times lower (ratio of half-lives). In case of alpha induced reactions no radioactive by-product is present, but the yield is low and highly enriched targets are required.

*Production of $^{134}$Ce/$^{134}$La (T$_{1/2}$ = 3.16 d / 6.45 min)*

The main possible light charged particle reaction routes for production of $^{134}$Ce include: $^{132}$Ba ($\alpha$,2n), $^{139}$La(p,6n), $^{138}$La(p,5n), $^{136}$Ce(p,x) and $^{136}$Ce(d,x) (Fig. 16).

In case of proton induced reaction on lanthanum long lived by-products are formed, while for the $^{132}$Ba($\alpha$,2n), $^{136}$Ce (p,2n) and $^{136}$Ce(d,3n) reactions we have only shorter-lived by-products, which can be totally or partially eliminated via appropriate cooling. Main problem is the low abundance of the $^{132}$Ba (0.101 %) and the $^{136}$Ce (0.19 %) in the natural target material.

*$^{135}$La (T$_{1/2}$ = 19.5 h) direct and indirect*

The excitation functions of the main light charged particle reaction routes for production of $^{135}$La directly or through the decay of $^{135}$Ce are shown in Fig. 17: $^{135}$Ba(p,n)$^{135}$La, $^{136}$Ba(p,2n)$^{135}$La, $^{134}$Ba(d,n)$^{135}$La, $^{135}$Ba(d,2n)$^{135}$La, $^{136}$Ce(p,x)$^{135}$La, $^{136}$Ce(d,x)$^{135}$La, $^{139}$La(p,xn)$^{135}$Ce-$^{135}$La, $^{138}$La(p,4n)$^{135}$Ce-$^{135}$La, $^{136}$Ce(p,x)$^{135}$Ce-$^{135}$La, $^{136}$Ce(d,x)$^{135}$Ce-$^{135}$La, $^{133}$Cs($\alpha$,2n)$^{135}$La and $^{132}$Ba($\alpha$,n)$^{135}$Ce.

The low energy reactions on enriched Ba and Cs isotopes provide $^{135}$La directly with high radionuclidic purity if the enery window is well controlled. Pure $^{135}$Ce can be produced only with the $^{132}$Ba($\alpha$,n) reaction, but all other reactions should also be taken into account, because the lanthanum decay product of the simultaneously produced cerium isotopes are either stable or have very long half-life.

*$^{131}$Cs (T$_{1/2}$ = 9.689 d) direct and indirect*

The main light charged particle reaction routes for production of $^{131}$Cs include: $^{131}$Xe(p,n)$^{131}$Cs, $^{130}$Xe(d,n)$^{131}$Cs, $^{132}$Ba(p,x)$^{131}$Cs, $^{130}$Ba(d,x)$^{131}$Cs, $^{133}$Cs(p,x)$^{131}$Ba-$^{131}$Cs, $^{133}$Cs(d,x)$^{131}$Ba-$^{131}$Cs, $^{129}$Xe($\alpha$,2n)$^{131}$Ba-$^{131}$Cs, $^{138}$La(p,x)$^{131}$Ba-$^{131}$Cs, $^{139}$La(p,x)$^{131}$Ba-$^{131}$Cs.

Low energy production routes $^{131}$Xe(p,n)$^{131}$Cs, $^{130}$Xe(d,n) $^{131}$Cs and $^{130}$Ba (d,x)$^{131}$Cs require highly enriched targets (Fig. 18). The target recovery is simpler in case of gas targets.

At higher energies the route through $^{131}$Ba is preferred. When relying on the $^{133}$Cs(p,3n) and $^{133}$Cs(d,4n) reactions no enriched target is required and the production cross section is high. Production by using lanthanum target is possible at high energy production machines.

## Summary and conclusion

The principal aim of this investigation was to measure the basic cross-section data for production of practically applicable radionuclides and to complement the database for development of theoretical models. In this study activation cross sections of the $^{nat}$La(p,xn)$^{139,137m,137g,135,134,133m}$Ce, $^{nat}$La(p,x)$^{135,134,133}$La , $^{133m,133g,131}$Ba nuclear reactions have been measured a up to 65 MeV, for the first time (except for $^{nat}$La(p,x)$^{139}$Ce). An acceptable agreement was found with the theoretical predictions in the TENDL libraries.

We shortly discuss the capability of the presently investigated reactions for routine production of some practically relevant radionuclides in comparison with other production routes. The proton induced nuclear reactions on lanthanum can be taken into account for production of $^{139}$Ce, $^{137m}$Ce and $^{135}$Ce/$^{135}$La. The main advantages are the high yield and the nearly monoisotopic $^{139}$La target.

*Acknowledgements*

*This work was done in the frame of MTA-FWO (Vlaanderen) research projects. The authors acknowledge the support of research projects and of their respective institutions in providing the materials and the facilities for this work.*

**Figures**

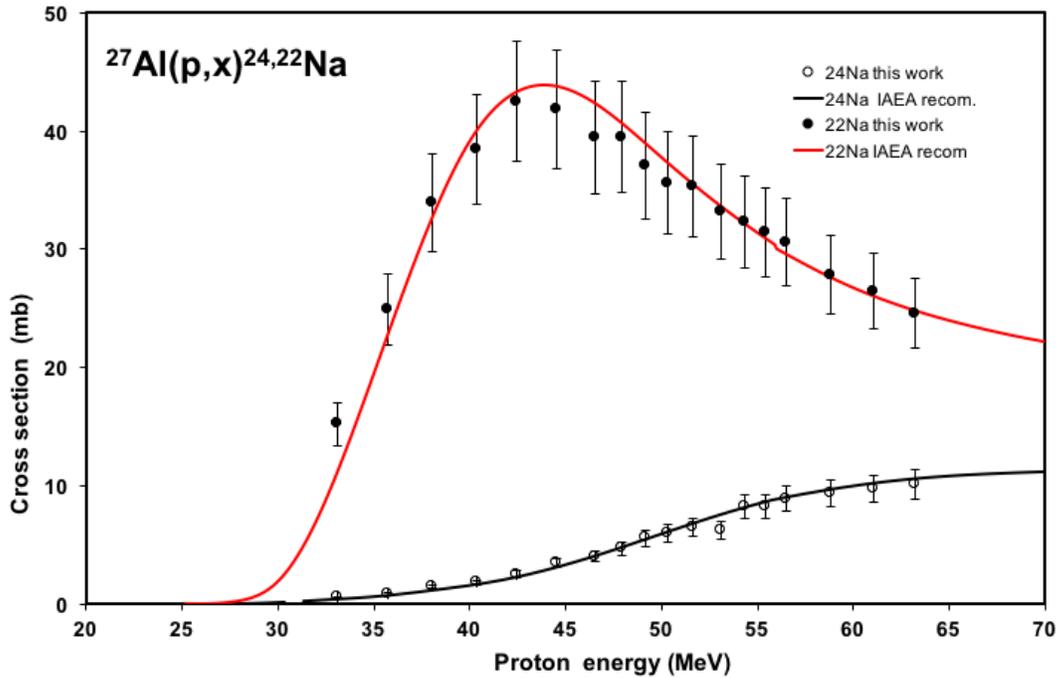

Fig. 1. Excitation function of the simultaneously irradiated $^{27}$Al(p,x)$^{24,22}$Na monitor reaction

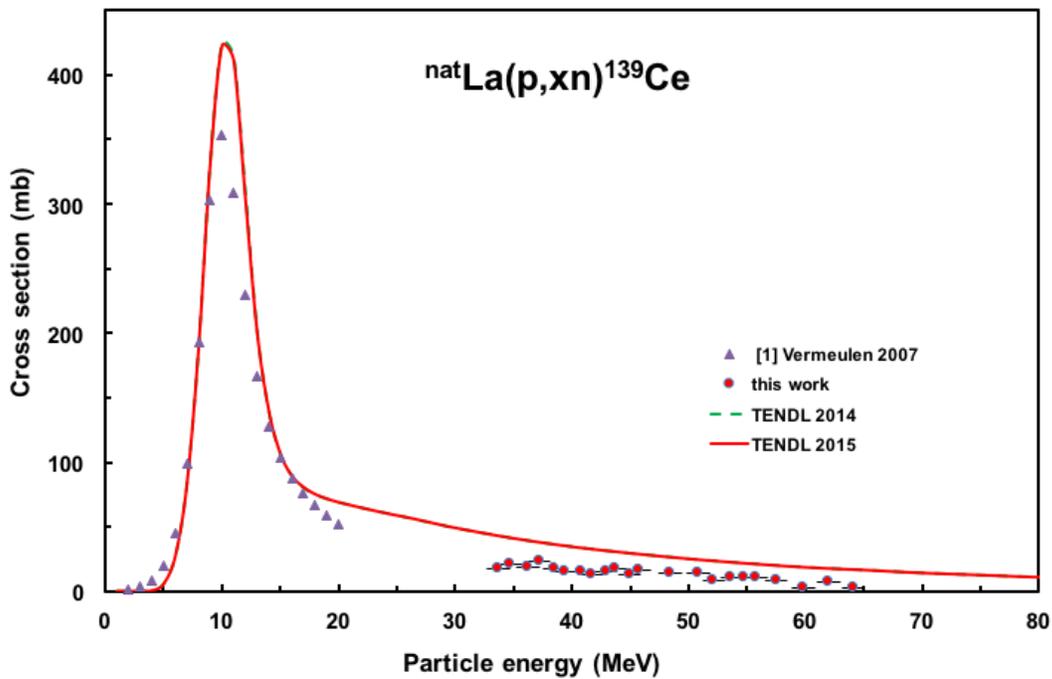

Fig. 2. Experimental excitation function for the $^{nat}$La(p,x)$^{139}$Ce reaction and comparison with TALYS theoretical code calculations

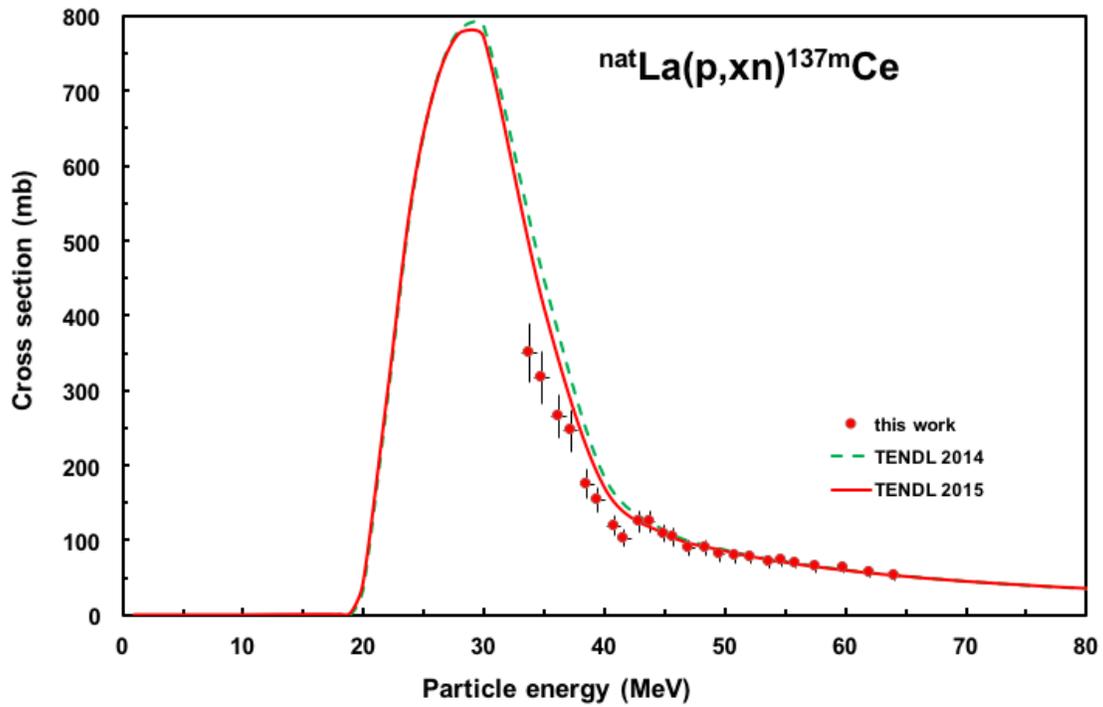

Fig. 3. Experimental excitation function for the $^{nat}$La(p,x)$^{137m}$Ce reaction and comparison with TALYS theoretical code calculations.

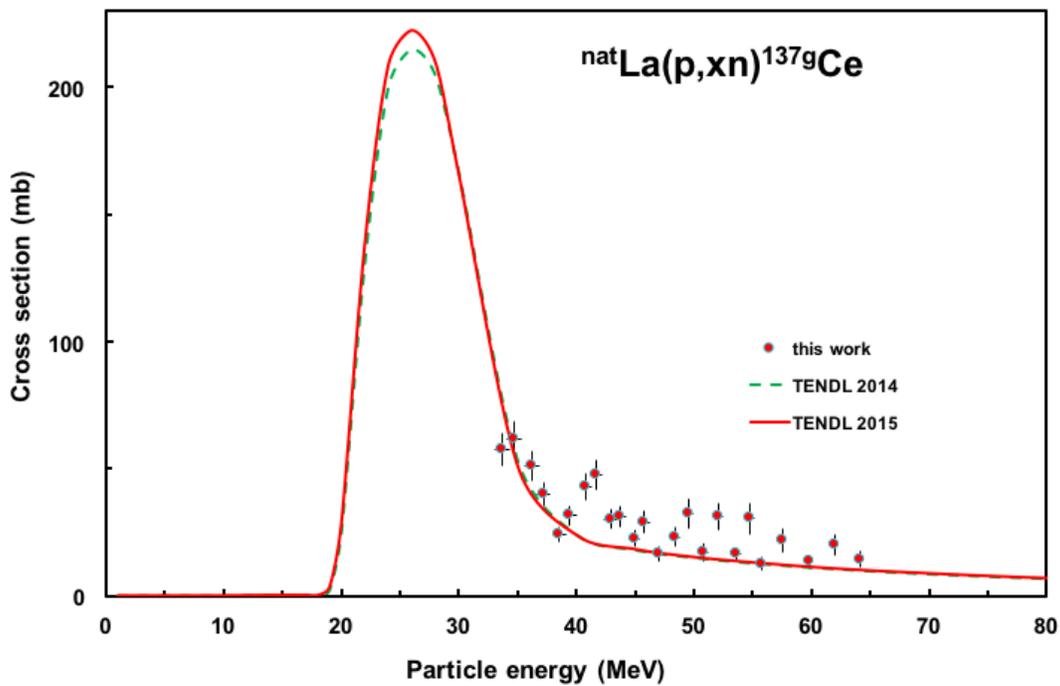

Fig. 4. Experimental excitation function for the $^{nat}$La(p,x)$^{137g}$Ce reaction and comparison with TALYS theoretical code calculations.

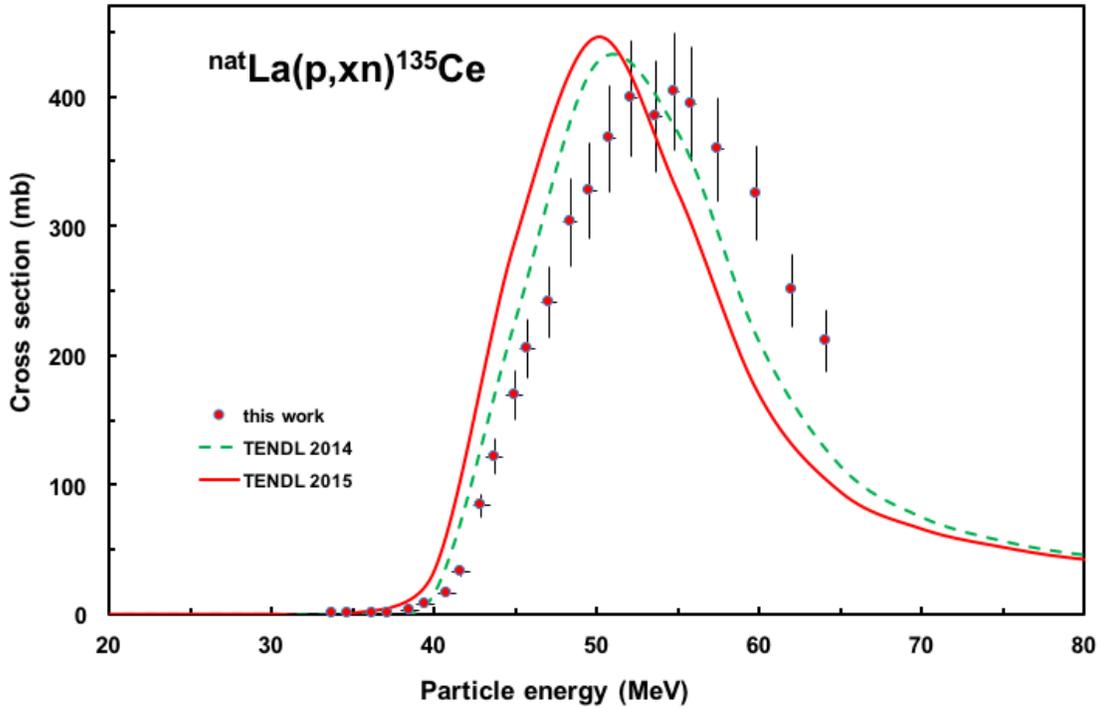

Fig. 5. Experimental excitation function for the $^{nat}$La(p,x)$^{135}$Ce reaction and comparison with TALYS theoretical code calculations.

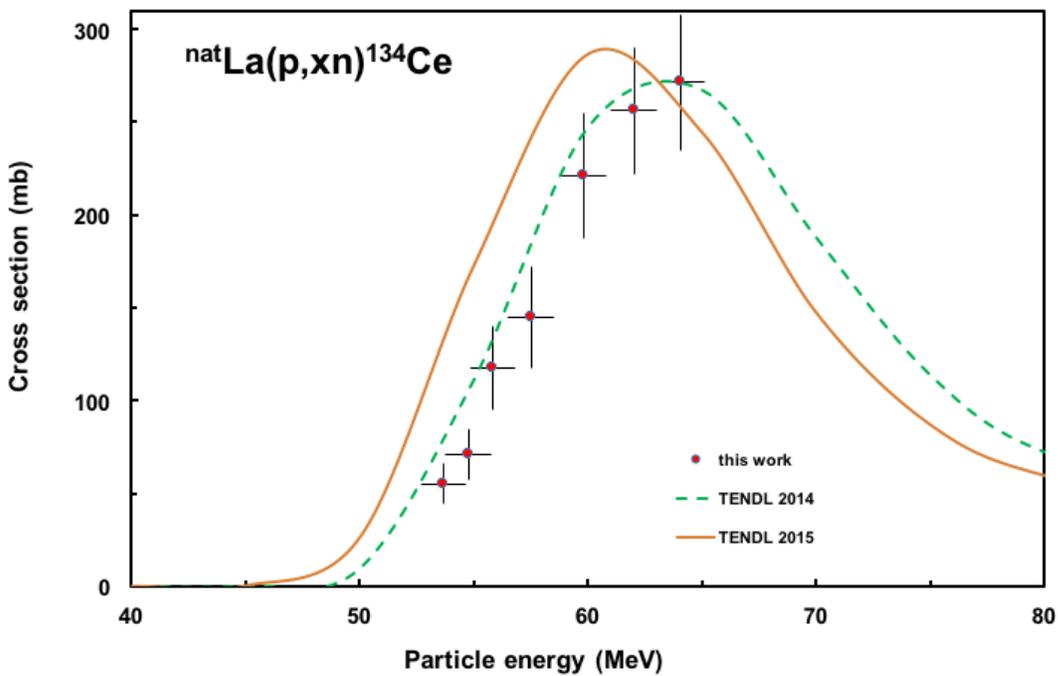

Fig. 6. Experimental excitation function for the $^{nat}$La(p,x)$^{134}$Ce reaction and comparison with TALYS theoretical code calculations.

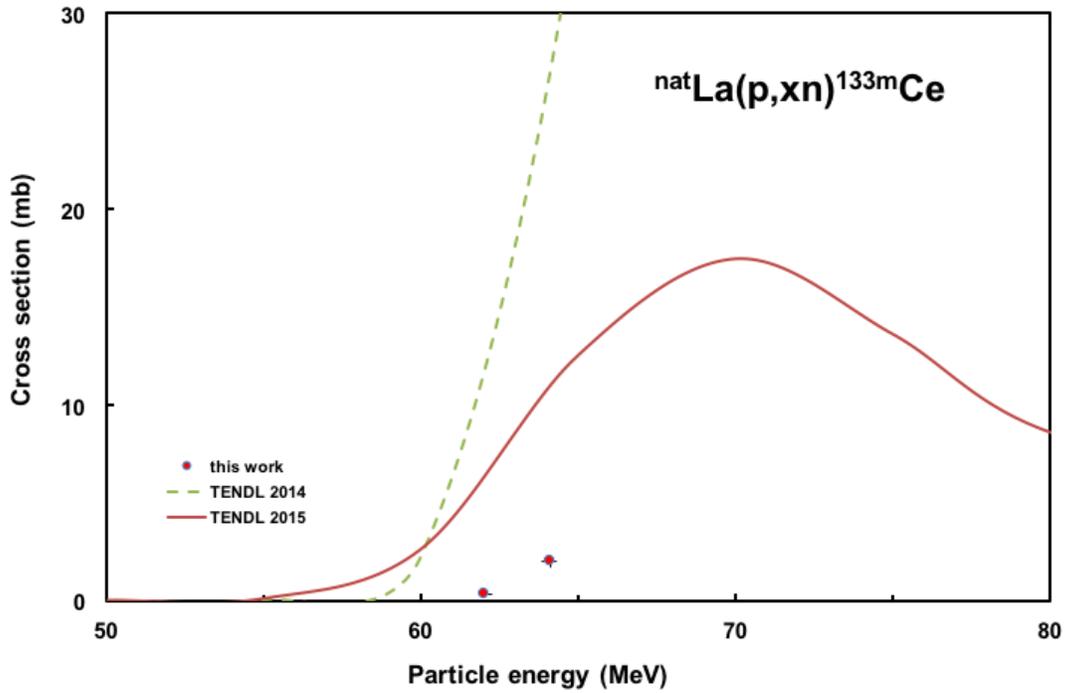

Fig. 7. Experimental excitation function for the $^{nat}$La(p,x)$^{133m}$Ce reaction and comparison with TALYS theoretical code calculations.

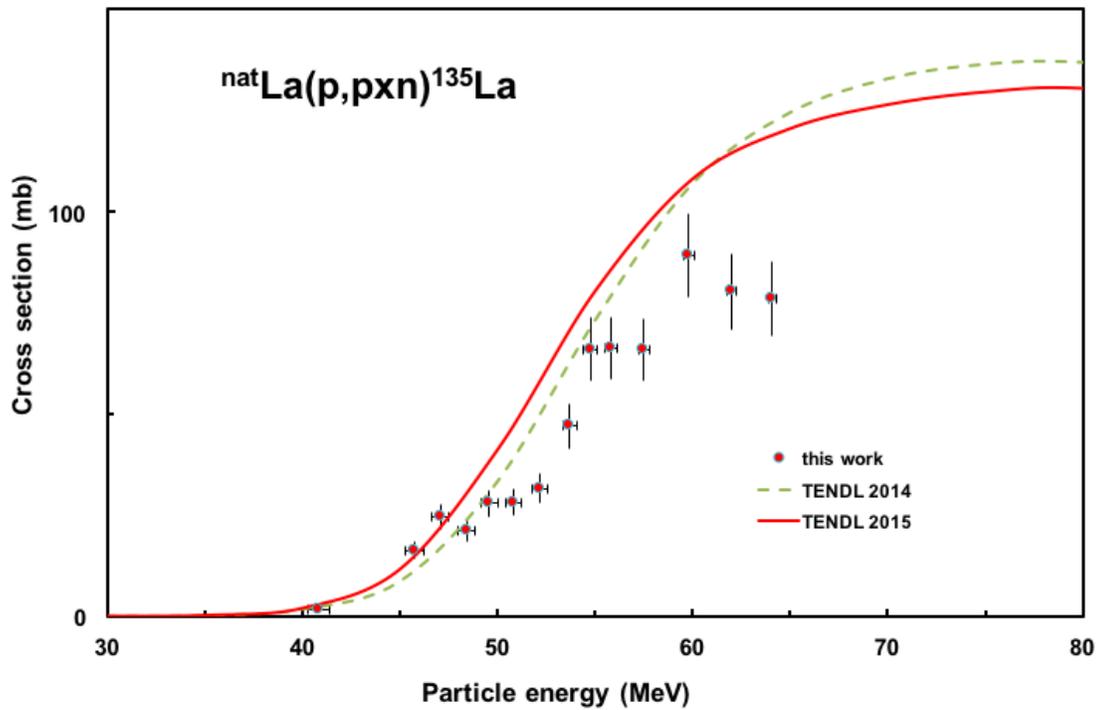

Fig. 8. Experimental excitation function for the $^{nat}$La(p,x)$^{135}$La reaction and comparison with TALYS theoretical code calculations.

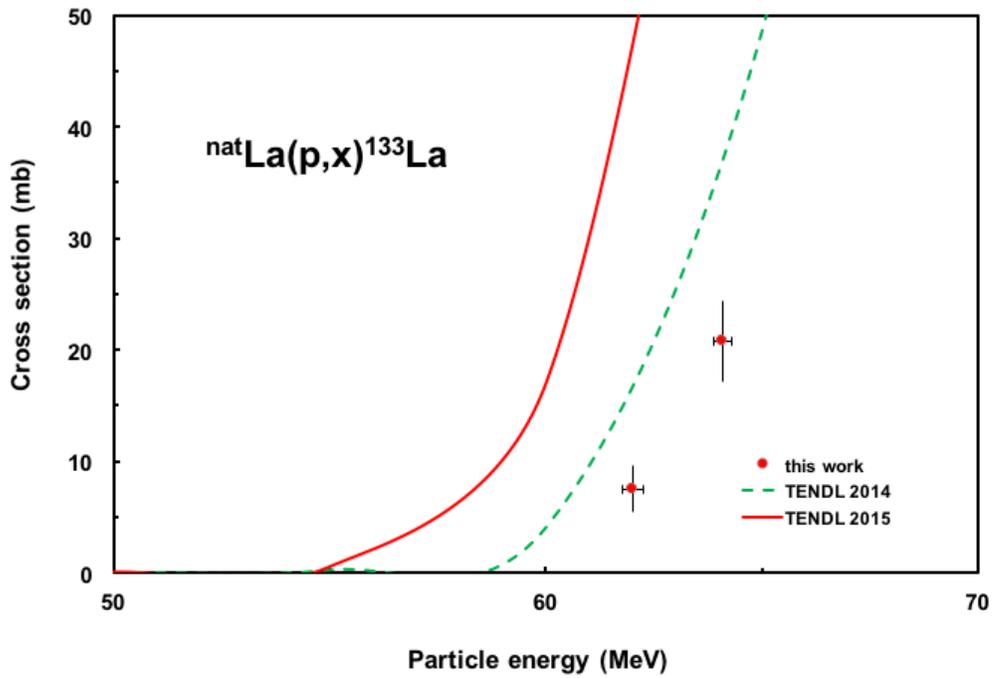

Fig. 9. Experimental excitation function for the $^{nat}$La(p,x)$^{133}$La reaction and comparison with TALYS theoretical code calculations.

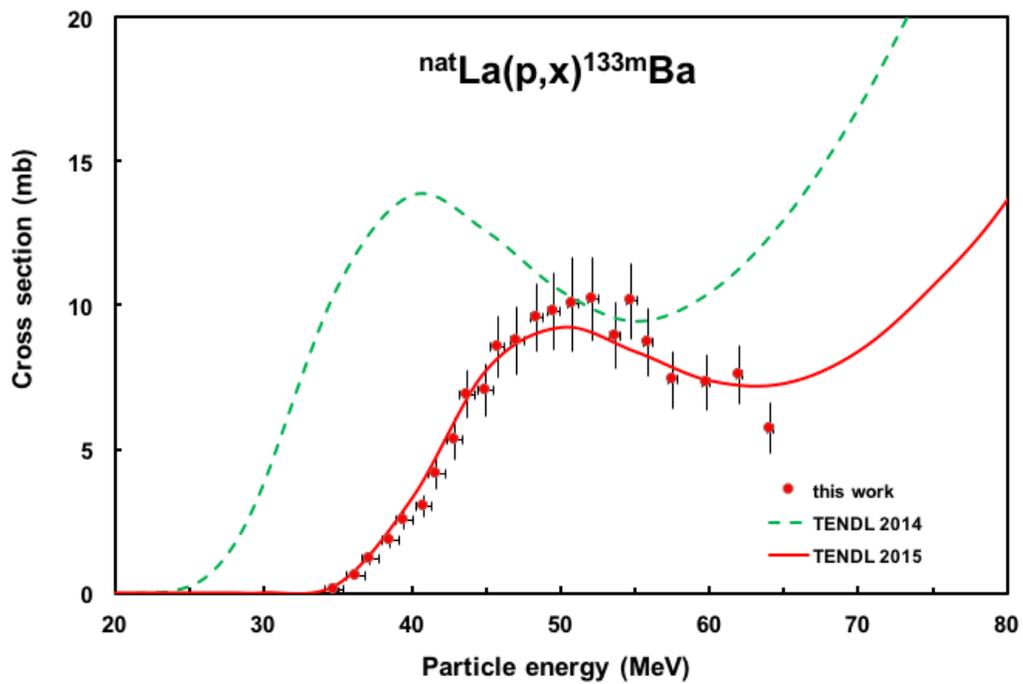

Fig. 10. Experimental excitation function for the $^{nat}$La(p,x)$^{133m}$Ba reaction and comparison with TALYS theoretical code calculations .

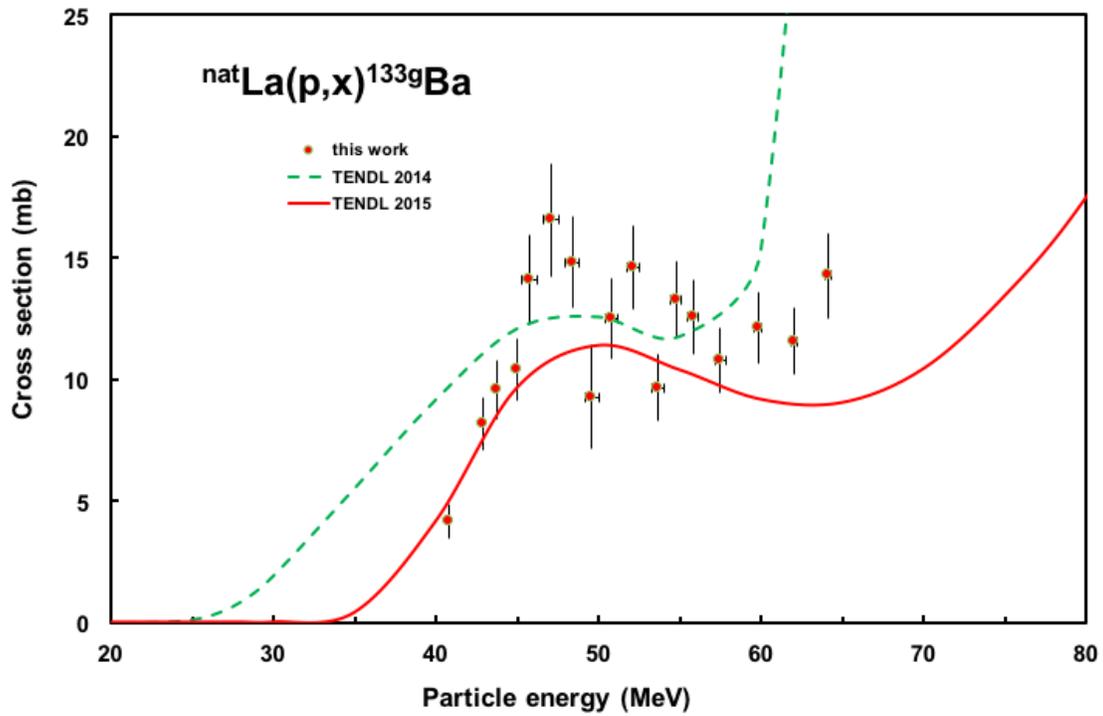

Fig. 11. Experimental excitation function for the $^{nat}$La(p,x)$^{133g}$Ba reaction and comparison with TALYS theoretical code calculations.

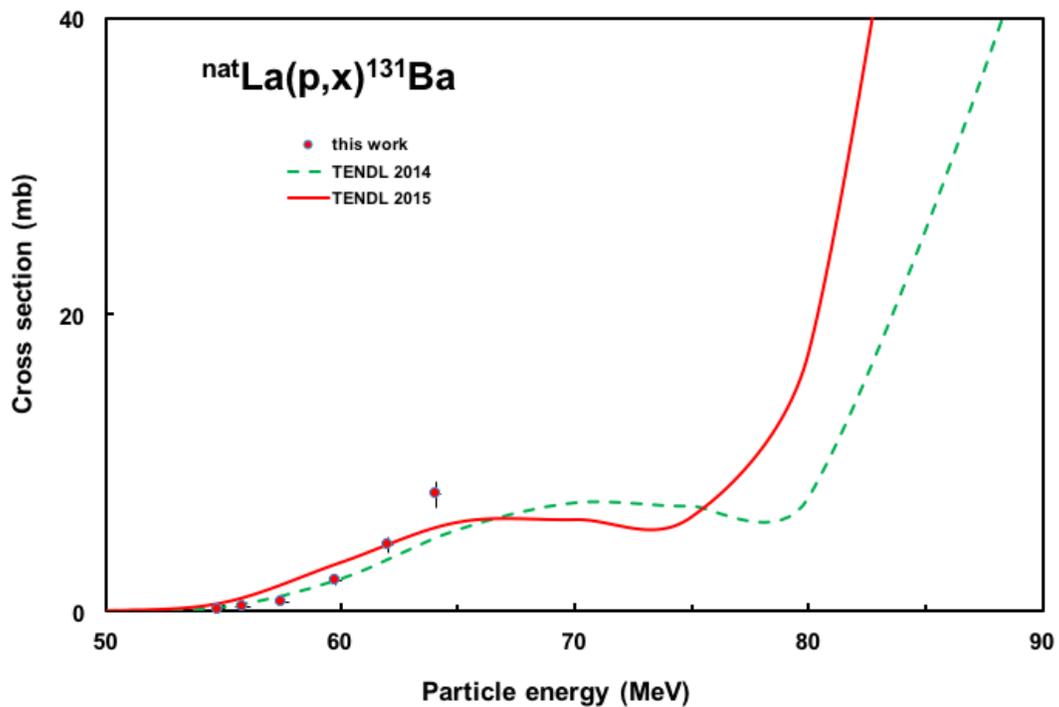

Fig. 12. Experimental excitation function for the $^{nat}$La(p,x)$^{131}$Ba reaction and comparison with TALYS theoretical code calculations.

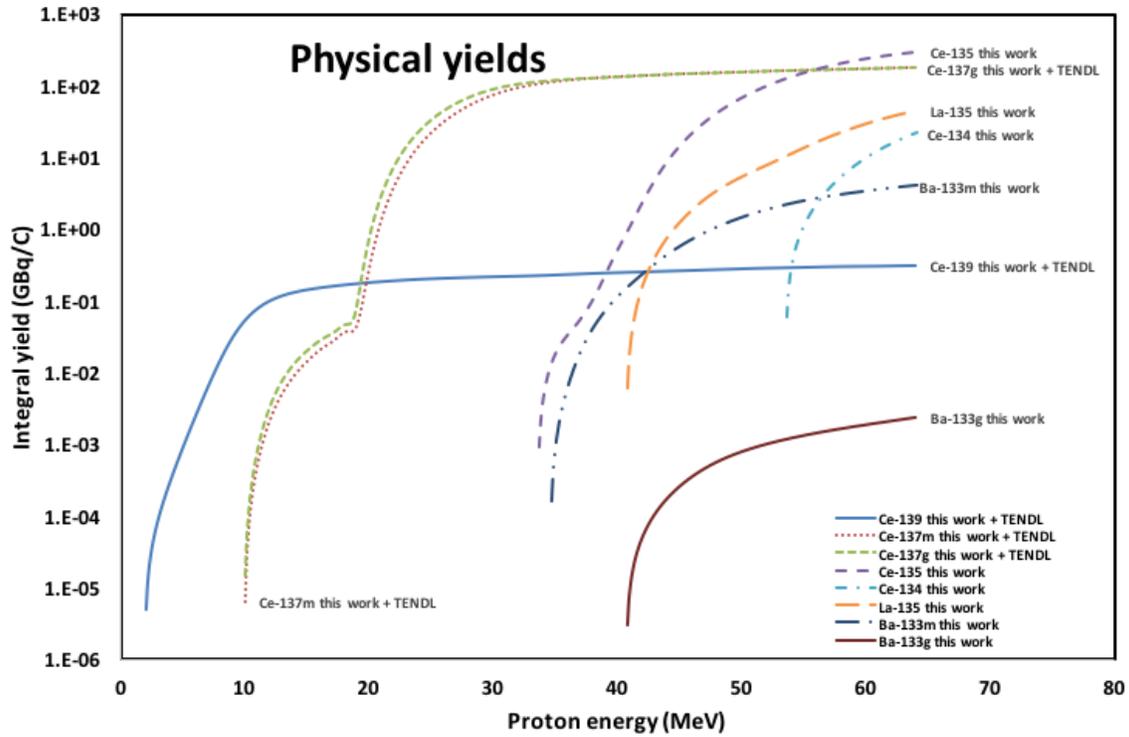

Fig. 13. Thick target yields for radionuclides of $^{nat}La(p, x)^{139, 137m,137g,135,134}Ce$, $^{nat}La(p,x)^{135}La$, $^{133m,133g}Ba$ nuclear reactions.

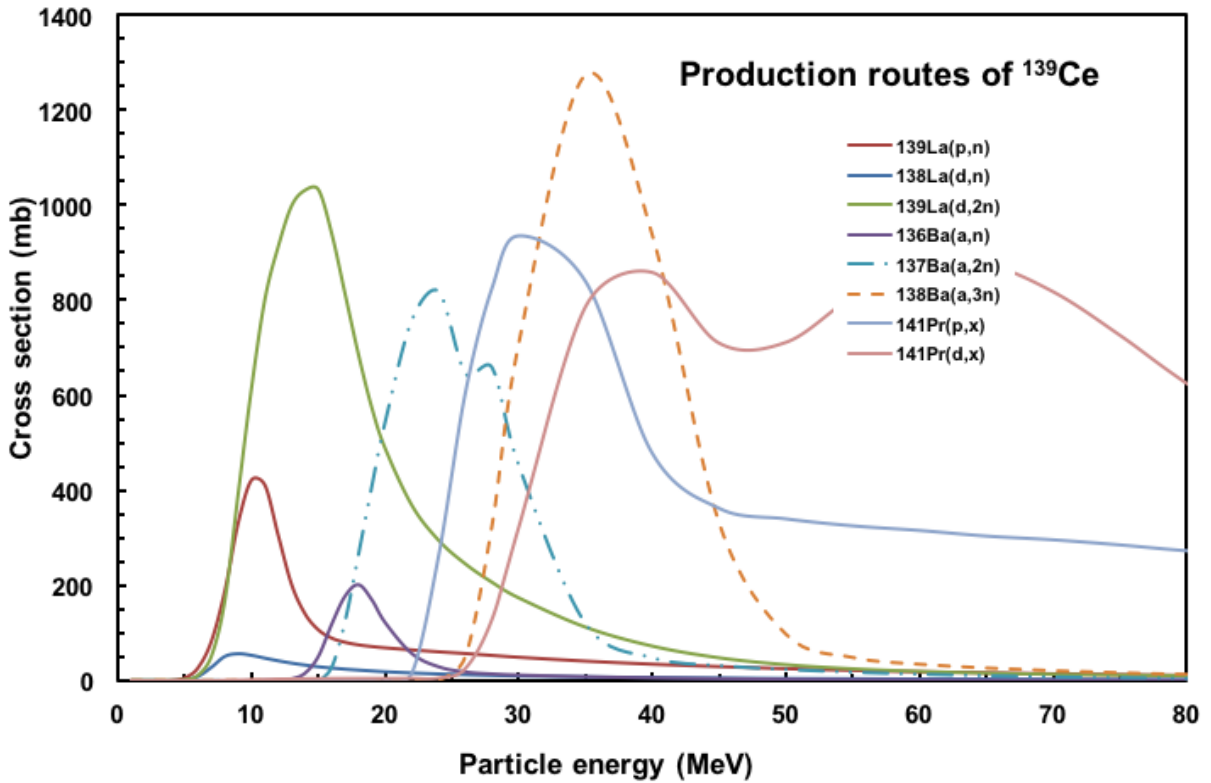

Fig. 14. Comparison of the production routes of $^{139}Ce$

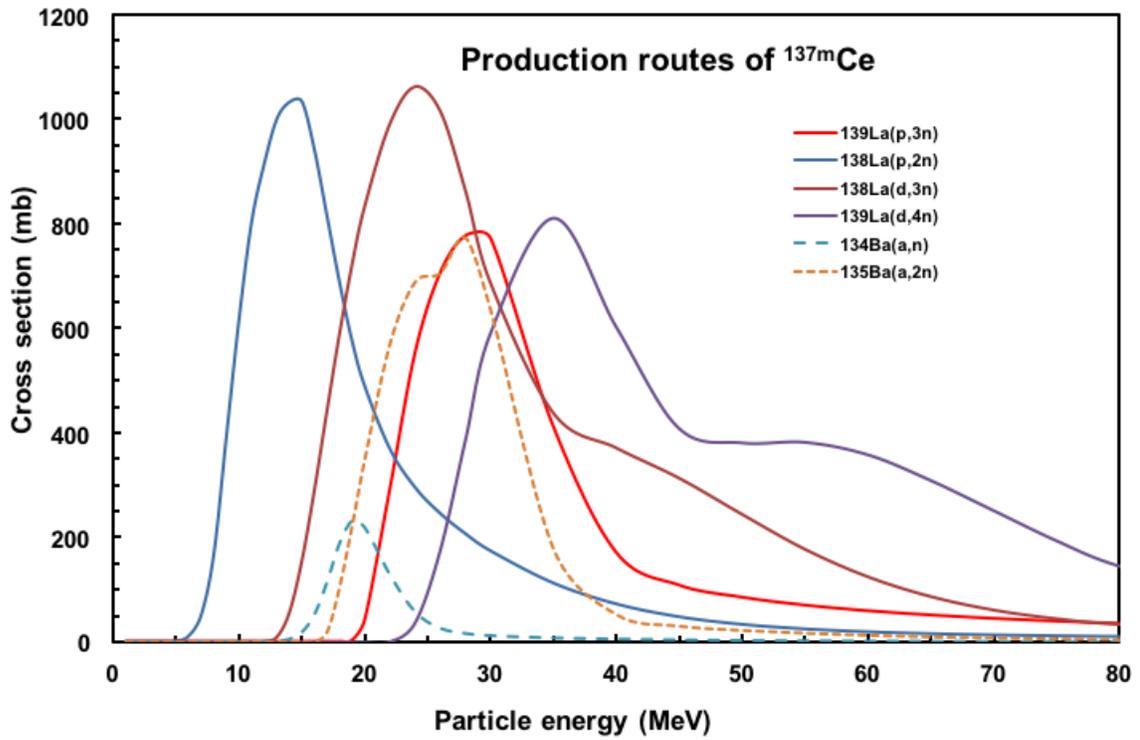

Fig. 15. Comparison of the production routes of $^{137m}$Ce

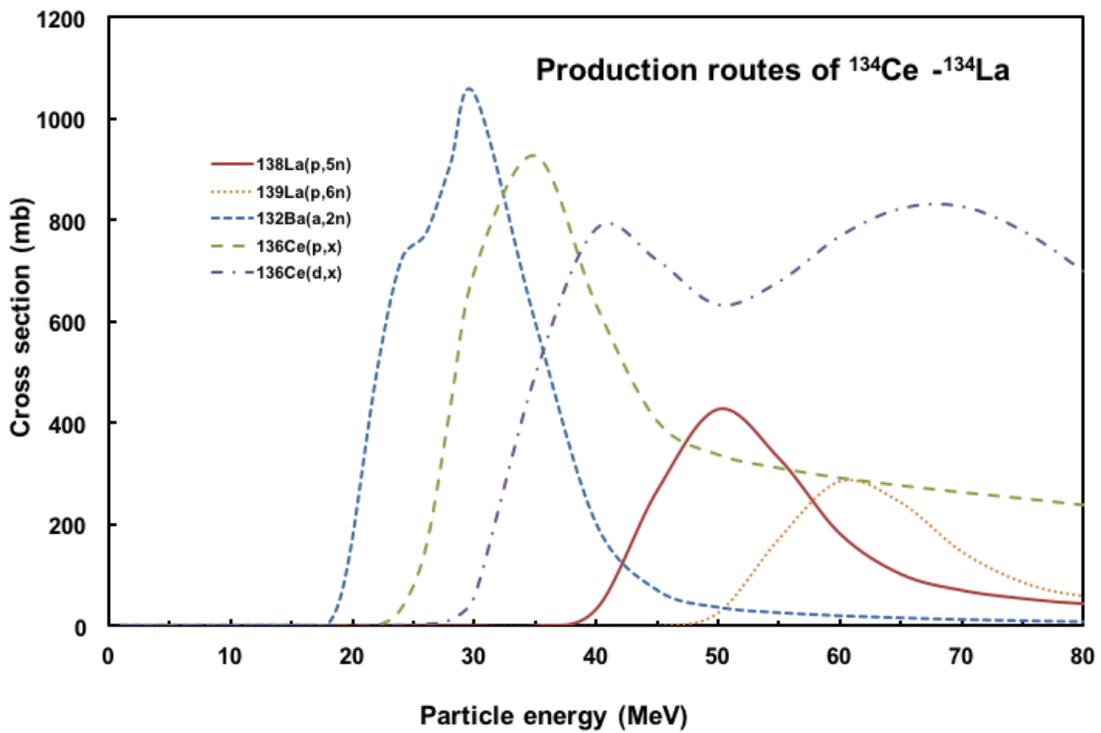

Fig. 16. Comparison of the production routes of $^{134}$Ce

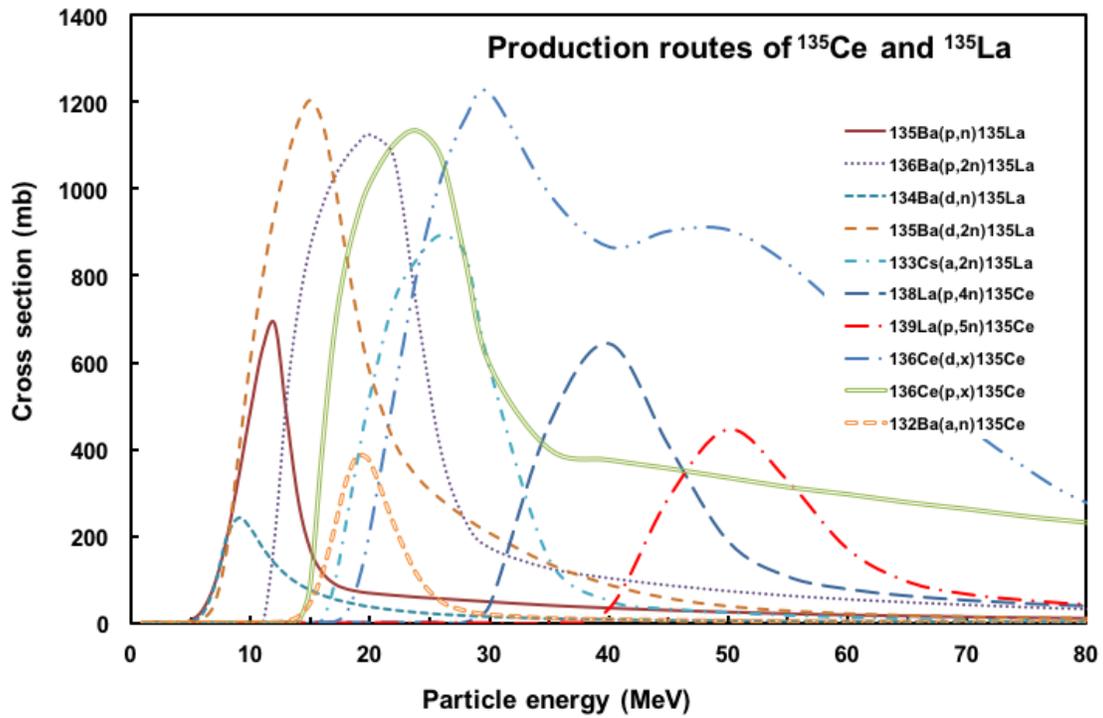

Fig. 17. Comparison of the production routes of the $^{135}$La and $^{135}$Ce

Fig. 18. Comparison of the production routes of $^{131}$Ba and $^{131}$Cs

*Activation cross-section measurement of deuteron induced reactions on cerium for biomedical applications and for development of reaction theory.*
Nuclear Instruments and Methods in Physics Research Section B: Beam Interactions with Materials and Atoms **316** :22-32

13 Tárkányi F. , Ditrói F. , Takács S. , Hermanne A. , Baba M. , Ignatyuk A. V.  (2016)
*Excitation functions for (d,x) reactions on $^{133}$Cs up to Ed=40 MeV.*
Applied Radiation and Isotopes **110** :109-117.

14 F. Tárkányi , S. Takács, F. Ditrói, A. Hermanne, A. V. Ignatyuk, Z. Szucs,
*Study of activation cross sections of deuteron induced reactions on barium*
Nuclear Instruments and Methods in Physics Research Section B: Beam Interactions with Materials and Atoms  2017(submitted)

15 Hermanne A., Adam-Rebeles R. , Tárkányi F., Takács S. , Csikai Gy. , Takács M. P. , Ignatyuk A. (2013)
*Deuteron induced reactions on Ho and La: Experimental excitation functions and comparison with code results.*
Nuclear Instruments and Methods in Physics Research Section B: Beam Interactions with Materials and Atoms **311:** 102-111.

16 Hermanne A. , Tárkányi F. , Takács S. , Ditrói F. (2016)
*Extension of excitation functions up to 50 MeV for activation products in deuteron irradiations of  Pr and Tm targets.*
Nuclear Instruments and Methods in Physics Research Section B: Beam Interactions with Materials and Atoms **383>** 81-88.

17 http://www.canberra.com/products/radiochemistry_lab/genie-2000-software.asp

18 Székely G.  (1985)
*FGM - a flexible gamma-spectrum analysis program for a small computer*
Computer Physics Communications **34:** 313-324

28